\documentclass[12pt]{article}
\usepackage{amssymb}
\begin{document}

\makeatletter
\def\@maketitle{\newpage
 \null
 {\normalsize \tt \begin{flushright} 
  \begin{tabular}[t]{l} \@date  
  \end{tabular}
 \end{flushright}}
 \begin{center} 
 \vskip 2em
 {\LARGE \@title \par} \vskip 1.5em {\large \lineskip .5em
 \begin{tabular}[t]{c}\@author 
 \end{tabular}\par} 
 \end{center}
 \par
 \vskip 1.5em} 
\makeatother
\topmargin=-1cm
\oddsidemargin=1.5cm
\evensidemargin=-.0cm
\textwidth=15.5cm
\textheight=22.5cm
\setlength{\baselineskip}{16pt}
\title{ Non-commutative Field Theory on  $S^4$ }
\author{  Ryuichi~{\sc Nakayama}\thanks{ nakayama@particle.sci.hokudai.ac.jp}
and Yusuke~{\sc Shimono}\thanks{yshimono@particle.sci.hokudai.ac.jp}
       \\[1cm]
{\small
    Division of Physics, Graduate School of Science,} \\
{\small
           Hokkaido University, Sapporo 060-0810, Japan}
}
\date{
  EPHOU-04-002  \\
hep-th/0404086 \\ 
April  2004  
}
%
%
\maketitle

\newcommand {\beq}{\begin{equation}}
\newcommand {\eeq}{\end{equation}}
\newcommand {\beqa}{\begin{eqnarray}}
\newcommand {\eeqa} {\end{eqnarray}}
\newcommand{\bm}[1]{\mbox{\boldmath $#1$}}
\newcommand{\Sq}{D(X)}
\newcommand{\al}{2\pi \alpha'}

 
\begin{abstract} 
In the previous paper (hep-th/0402010) we proposed a matrix configuration for 
a non-commutative $S^4$ (NC4S) and constructed a non-commutative (star) 
product for field theories on NC4S. In the present paper we will show that 
any matrix can be expanded in terms of the matrix configuration representing 
NC4S just like any matrix can be expanded into symmetrized products of the 
matrix configuration for non-commutative $S^2$. Then a scalar field theory 
on NC4S is constructed. Our matrix configuration describes two $S^4$'s joined 
at the circle and the Matrix theory action contains a projection matrix 
inside the trace to restrict the space of matrices to that for one $S^4$.
\end{abstract}
\newpage
\setlength{\baselineskip}{18pt}

\section{Introduction}
\hspace{5mm}
The construction of the non-commutative $S^4$ (NC4S) has not been 
straightforward. In \cite{CLT} matrix configuration for NC4S was proposed 
to describe spherical Longitudinal 5-branes (L5-branes) in the context of 
the Matrix Theory.\cite{BFSS}\cite{IIB} NC4S was also considered to 
construct an 
example of finite 4d field theory.\cite{GKP}\cite{Oconner}
There is, however, a problem in describing the 
fluctuating L5-branes and it was concluded that `fuzzy $S^4$' is 6 
dimensional, {\it i.e.}, a fuzzy $S^2$ fibre bundle over a non-associative 
$S^4$.\cite{HoRangoolam} \cite{Rangoolam}

In the above works it was assumed that the non-commutative $S^4$ algebra is 
covariant under $SO(5)$. In the flat, non-compact case, however, the Moyal 
space is not invariant under space rotations except for two dimensions. 
The Moyal algebra
\begin{equation}
\ [ x^{\mu}, x^{\nu} ]= i \theta^{\mu \nu}, \qquad (\mu, \nu=1,2,..,d)
\end{equation}
is not covariant under $SO(d)$ but only under $SO(2) \times SO(2) \cdots$.
Let us suppose that there exists an $SO(5)$ covariant non-commutative algebra
on $S^4$. In the large radius limit this algebra is expected to be replaced by
a non-commutative algebra for the tangent space around some point on 
$S^4$. This algebra would be covariant under $SO(4)$ which is the symmetry 
around the axis connecting the point and the origin.  This contradicts with
the non-existence of the $SO(4)$ covariant non-commutative algebra in
the flat, non-compact space.  Therefore we must abandon the $SO(5)$ covariant
formulation.

In the previous paper\cite{NS} the present authors proposed a matrix 
configuration $\hat{X}_0^A$ ($A=1,..,5$) for non-commutative $S^4$ by means of 
the tensor products of two $SU(2)$ generators, each in the spin $j_1$ 
and $j_2$ 
representations, respectively. They satisfy the constraint of the sphere. 
\begin{equation}
(\hat{X}_0^A)^2=\hat{R}^2 \ \bm{1}
\label{S4constraint}
\end{equation}
The algebra of these matrices is not a Lie algebra and not covariant under 
$SO(5)$.

The motivation of the above construction was that the non-commutative algebra 
of $S^2$ is identified as $SU(2)$ algebra\cite{Madore} and the correspondence 
between the generators $T^a$ of $SU(2)$ and the non-commutative coordinates 
$x^a$ is established\cite{deWitt}; matrices are mapped onto functions on 
(commutative) $S^2$ and the matrix multiplication is mapped onto a 
non-commutative product (star product) of the functions. When the 
representation of $SU(2)$ is that with spin $j$, any matrix of size 
$N=2j+1$ can be expanded into symmetrized products of the generators $T^a$ 
up to order $2j$. Correspondingly, the functions on $S^2$ are spanned 
by polynomials of $x^a$ up to order $2j$. Then the matrices constructed by 
tensor products of two matrices will be naturally mapped onto functions 
on a four dimensional manifold and the non-commutative product for 
this manifold will be obtained by tensor product of those 
for two $S^2$'s. Because the matrix configuration satisfies the constraint of 
$S^4$, (\ref{S4constraint}), it is expected that non-commutative field theory 
on $S^4$ will be obtained. 

The functions on $S^4$ and the non-commutative product constructed in 
\cite{NS} have singularity on a circle ($(X^4)^2+(X^5)^2=R^2$) on $S^4$, 
which is an ambiguity 
(indeterminateness) of the values of these functions and the product. 
This is due to the $S^2 \times S^2$ parametrization of $S^4$ which is used 
in the construction of the star product. This singularity is so mild that 
this is not a serious problem in describing non-commutative field theories 
on $S^4$; in non-commutative field theories differentiation is
a star-commutator and differentiation does not create more singular terms.  
In the description of a non-commutative 
field theory in terms of the Matrix theory there are no singularities, 
because finite dimensional matrices and traces of them are well-defined. 

We will show that any matrix of size $N=(2j_1+1)(2j_2+1)$ can be 
expressed in terms of the matrix $S^4$ configuration $\hat{X}_0^A$, if and
only if $j_1$ is half an odd integer. In contrast to the $S^2$ case this is 
not a polynomial of $\hat{X}_0^A$'s. Instead, a matrix is expanded into 
products of the functions of $\hat{X}_0^A$ which are in exact correspondence 
with the functions on $S^4$ given by the Matrix $\leftrightarrow$ Function 
correspondence. 

Then we will construct a scalar field theory on NC4S. The Matrix theory
action contains a projection matrix inside the trace. This is because our  
parametrization of $S^4$ actually describes two $S^4$'s
and we must restrict the space of the matrices into the subspace describing 
only one $S^4$. The cyclic property of the trace is partially broken, but 
this is not an obstacle for writing the Matrix theory action for the 
scalar theory. In the case of the gauge theory there will be a restriction on 
the gaue symmetry. 

Organization of the paper is as follows.  In sec.2 we will give some 
properties of the matrix configuration $\hat{X}^A_0$ for $S^4$. 
Especially, we show that any matrix can be expanded in terms of the base 
matrices constructed from the matrix configuration $\hat{X}^A_0$.  
In sec.3 we will construct a scalar thery on NC4S. 
There is some summary in sec.4.

\vspace*{1cm}

Finally a comment on the notation: 
throughout this paper we will use the initial lowercase romans 
$a$, $b$, $c,...$ for values $1,2,3$ and the middle romans $i,j,..=
4,5$, while the capitals $A,B,..$ will run from 1 to 5. 

\section{Matrix configuration}
\hspace*{5mm}
In \cite{NS} we proposed the following matrix configuration for NC4S.
\begin{eqnarray}
\hat{X}_0^a &=&\frac{\alpha}{j_1(j_1+1)} \ T_{(j_1)}^3 \otimes T_{(j_2)}^a, 
\qquad (a=1,2,3) \nonumber \\
\hat{X}_0^4&=& \frac{\beta}{2j_1(j_1+1)} \ T_{(j_1)}^1 \otimes 
\bm{1}_{2j_2+1}, \qquad \hat{X}_0^5= \frac{\beta}{2j_1(j_1+1)} 
\ T_{(j_1)}^2 \otimes \bm{1}_{2j_2+1}. 
\label{matrixconfiguration}
\end{eqnarray}
Here $T_{(j)}^a$ $(a=1,2,3)$ are the generators of the spin-$j$ 
representation of $SO(3)$ and satisfy 
\begin{equation}
\ [ T_{(j)}^a, T_{(j)}^b ]= i \epsilon_{abc} \ T_{(j)}^c
\end{equation}
$\alpha$ and $\beta$ are normalization constants.\footnote{ Actually, we 
mainly discussed only the case of $j_1=1/2$ in \cite{NS} and in sec.6 of 
that paper we made a brief comment on general $j_1$.}

This configuration has the following properties.\footnote{This 
configuration actually describes two $S^4$'s joined at a circle. To obtain 
non-commutative field theories on $S^4$ we must perform a projection onto  
only one $S^4$. (sec.3)}
\begin{description}
\item [(1)] The matrices (\ref{matrixconfiguration}) satisfies the matrix 
analogue of the equation for a four-dimensional ellipsoid.
\begin{equation}
\left( \hat{X}_0^a \right)^2 + 4\frac{\alpha^2}{\beta^2}j_2(j_2+1) \ 
\left( \hat{X}_0^i \right)^2 = \hat{R}^2 \ \bm{1}
\label{ellipsoid}
\end{equation}
Here $\hat{R}$ is a constant defined by 
\begin{equation}
\hat{R} = \alpha \  \sqrt{\frac{j_2(j_2+1)}{j_1(j_1+1)}}.
\label{Rhat}
\end{equation}
If we choose 
\begin{equation}
\beta=2 \alpha \ \sqrt{j_2(j_2+1)}, 
\label{betaalpha}
\end{equation}
eq (\ref{ellipsoid}) becomes eq (\ref{S4constraint}), the constraint of $S^4$. 

\item [(2)] This configuration satisfies the following algebra.\footnote{
This algebra is new and defferent from the one we gave in \cite{NS}.} 
\begin{eqnarray}
&& [ \hat{X}_0^a, \hat{X}_0^b] = \frac{2\alpha}{\beta^2} \ j_1(j_1+1) \ 
\epsilon_{abc} \ \left( \hat{X}_0^c \ [\hat{X}_0^4,\hat{X}_0^5]+ [\hat{X}_0^4,
\hat{X}_0^5] \ \hat{X}_0^c \right), \nonumber \\
&& [ \hat{X}_0^a, \hat{X}_0^i] = \frac{j_1(j_1+1)}{\alpha} \ 
\epsilon_{abc} \epsilon_{ij} \ \left( \hat{X}_0^b \ [\hat{X}_0^c,\hat{X}_0^j]
+ [\hat{X}_0^c,\hat{X}_0^j] \ \hat{X}_0^b \right), \nonumber \\
&& (\hat{X}_0^a)^2 \ [ \hat{X}_0^4,\hat{X}_0^5 ]+ [ \hat{X}_0^4,\hat{X}_0^5 ] 
\ (\hat{X}_0^a)^2 = \frac{\beta^2}{2\alpha \ j_1(j_1+1)} \ 
 \epsilon_{abc} \ \hat{X}_0^a \ \hat{X}_0^b \ \hat{X}_0^c \nonumber \\
&& \label{S4algebra}
\end{eqnarray}
This is not covariant under $SO(5)$ but only under $SO(3) \times SO(2)$.

Properties (1) and (2) can be checked by explicit calculation.

\item [(3)] Any $N \times N$ matrix $\hat{M}$ ($N=(2j_1+1)(2j_2+1)$) can be 
expressed in terms of $\hat{X}_0^A$, if $j_1$ is half an odd integer.
This is analogous to the cases of non-commutative torus\cite{SeibergWitten} 
and non-commutative $S^2$\cite{deWitt}. In the case of non-commutative torus 
any matrix can be written as a linear combination of $P^n Q^m$, where $P$, 
$Q$ are matrices satisfying $PQP^{-1}Q^{-1}= \exp (2\pi i/N) \bm{1}$. In the 
case of non-commutative $S^2$ any matrix can be expanded into symmetrized 
products of $T^a$, the generators of $SO(3)$. 

Let us note that any $2j+1 \times 2j+1$ matrix can be expanded in terms of the 
products of $T^a_{(j)}$'s. Therefore it is sufficient to express 
$T^a_{(j_1)} \otimes \bm{1}$ and $\bm{1} \otimes T^a_{(j_2)}$ in terms of 
$\hat{X}_0^A$.  First of all from the last two eqs of 
(\ref{matrixconfiguration}) we obtain
\begin{equation}
T^i_{(j_1)} \otimes \bm{1} = \frac{2j_1(j_1+1)}{\beta} \ \hat{X}_0^{i+3}
\qquad (i=1,2).
\end{equation}
We next compute 
\begin{equation}
(\hat{X}_0^a)^2 = \left(\frac{\alpha}{j_1(j_1+1)}\right)^2 \ (T^3_{(j_1)})^2
\otimes (T^a_{(j_2)})^2 = \frac{\alpha^2 j_2(j_2+1)}{(j_1(j_1+1))^2}  
\ (T^3_{(j_1)})^2 \otimes \bm{1}.
\label{T31}
\end{equation}
By taking the square root of both sides we have
\begin{equation}
T^3_{(j_1)} \otimes \bm{1} = \frac{j_1(j_1+1)}{\alpha \sqrt{j_2(j_2+1)}} \ 
\hat{D}. \label{T3}
\end{equation}
Here the matrix $\hat{D}$ is defined by
\begin{equation}
\hat{D} \equiv \left( (\hat{X}_0^a)^2 \right)^{1/2}. 
\label{D}
\end{equation}
The sign convention of the square root must be specified. 
We note that the non-vanishing eigenvalues $\lambda$ of (\ref{T31}) are all 
doubled. The square root is then defined such that both signs, 
$\sqrt{\lambda}$ and $-\sqrt{\lambda}$, are all included in the 
eigenvalues of (\ref{T3}) and (\ref{D}). 

To obtain $\bm{1} \otimes T_{(j_2)}^a$ we must divide $\hat{X}_0^a$ by 
$T^3_{(j_1)} \otimes \bm{1}$ and the latter matrix should not have the 
eigenvalue $0$. Hence $j_1$ must be half an odd integer.
In this case we obtain
\begin{equation}
\bm{1} \otimes T^a_{(j_2)} = \sqrt{j_2(j_2+1)} \ \hat{X}_0^a \ \hat{D}^{-1}.
\end{equation}

To conclude, any $N \times N$ matrix $\hat{M}$ can be expanded in terms of the 
products of the matrices, 
\begin{equation}
\hat{X}_0^4, \quad \hat{X}_0^5, \quad \hat{D}, \quad \hat{X}_0^a \ 
\hat{D}^{-1}\ \quad (a=1,2,3), 
\label{BaseMatrix}
\end{equation}
as long as $j_1$ is half an odd integer. Among these there are matrices which 
are not polynomials of $\hat{X}_0^A$. 

\end{description}

\section{Scalar field theory on NC4S}
\hspace*{5mm}
\subsection{Functions and star product on $S^4$}
\hspace*{5mm}
In the previous paper \cite{NS} we introduced the Matrix $\leftrightarrow $
Function correspondence on the NC4S in terms of the similar
correspondence on $S^2$. In this subsection we will recall the results 
of \cite{NS}. 

We introduced the following correspondence.\footnote{In \cite{NS} we 
considered only the case $j_1=1/2$. }
\begin{equation}
T^a_{(j_1)} \leftrightarrow j_1 \ \frac{x^a}{r}, \qquad 
T^a_{(j_2)} \leftrightarrow j_2 \ \frac{y^a}{\rho}.
\label{TxTy}
\end{equation}
Here $x^a$ and $y^a$ are coordinates on the two $S^2$'s and 
$r=\sqrt{(x^a)^2}$, $\rho=\sqrt{(y^a)^2}$. Then the configuration 
(\ref{matrixconfiguration}) is realized by the following coordinates
\begin{eqnarray}
X^a &=& \frac{\alpha j_2}{j_1+1}  \ \frac{x^3}{r} \ \frac{y^a}{\rho}, \quad 
(a=1,2,3) \nonumber \\
X^4 &=& \frac{\beta}{2(j_1+1)}  \ \frac{x^1}{r}, \qquad 
X^5 = \frac{\beta}{2(j_1+1)}  \ \frac{x^2}{r}
\label{Xxy}
\end{eqnarray}
Eqs (\ref{TxTy}), (\ref{Xxy}) constitute {\em the Matrix $\leftrightarrow $
Function correspondence}. 

One finds that
\begin{equation}
\frac{(j_1+1)^2}{(\alpha j_2)^2} \ (X^a)^2+ \frac{4(j_1+1)^2}{\beta^2} \
(X^i)^2=1. 
\end{equation}
For
\begin{equation}
\alpha=\alpha_{\ast} \equiv \frac{j_1+1}{j_2}\ R, 
\qquad \beta=\beta_{\ast} \equiv 2(j_1+1)\ R,
\label{roundS4}
\end{equation}
where $R$ denote the radius of the commutative $S^4$, 
this gives a constraint equation for $S^4$. Note that these constants do not 
satisfy (\ref{betaalpha}). This is interpreted as a quantum effect.  
The discrepancy disappears only in the limit $j_1, j_2 \rightarrow \infty$. 
In this limit the matrix algebra (\ref{S4algebra}) becomes commutative and 
the limiting geometry is expected to be that of a commutative $S^4$. 
In what follows we will choose these values (\ref{roundS4}) of 
$\alpha$, $\beta$.

Matrix multiplication is realized on the functions on $S^4$ by a 
non-commutative product $\star$\cite{NS}, which is induced by the 
non-commutative products $\ast_x$, $\ast_y$ on $S^2$ 
\cite{Presnejder} \cite{HNT} \cite{Alekseev} \cite{Matsubara},   
\begin{eqnarray}
f(x) \ast_x g(x)
& =& f(x) \ g(x)  +\sum_{m=1}^{2j_1} \lambda_1^m \ C_m(\lambda_1)
J_{a_1 b_1}(x) \cdots J_{a_m b_m}(x) \nonumber \\
&& \qquad  \qquad \quad \qquad \times \partial_{a_1} \cdots \partial_{a_m} 
\ f(x) \ \partial_{b_1} \cdots \partial_{b_m} \ g(x) 
\label{starS2}
\end{eqnarray}
and the similar eq for $f(y) \ast_y g(y)$. Here $\lambda_1=1/2j_1$. 
$f(x)$ and $g(x)$ are functions of $x^a/r$. $(r=\sqrt{(x^a)^2})$  
$C_m(\lambda)$ and $J_{ab}(x)$ are defined by
\begin{equation}
C_m(\lambda)=\frac{\lambda^m}{m! (1-\lambda)(1-2\lambda) 
\cdots (1-(m-1)\lambda)}, 
\end{equation}
\begin{equation}
J_{ab}(x)=r^2 \delta_{ab}-x^a \ x^b +i \ r \epsilon_{abc} \ x^c.
\end{equation}
In terms of two $\ast$'s for $x^a$ and $y^a$ we can define a non-commutative 
product on $S^4$.
\begin{equation}
F(X) \star G(X) \equiv F(X) \ \ast_x \otimes \ast_y \ G(X)
\label{starS4}
\end{equation}
Here $F(X)$ and $G(X)$ must be regarded as functions of $x^a$ and $y^a$: 
the factor $\ast_x$ of $ \ast_x \otimes \ast_y$ operates on $x^a$ and $\ast_y$ 
acts on $y^a$, respectively. The final expression must be re-expressed in 
terms of the $S^4$ coordinate, $X^A$. The explicit expression of $\star$ 
for $j_1=j_2=1/2$ representation was worked out in \cite{NS}.

The functions $F(X)$ on $S^4$ are the products of those on the two $S^2$'s.
They are polynomials of $x^a$ up to order $2j_1$ and those of $y^a$ up to
order $2j_2$, respectively. 
\begin{eqnarray}
F(X)&=& \sum_{m=0}^{2j_1} \ \sum_{n=0}^{2j_2} \ 
W_{a_1\ldots a_m;b_1 \ldots b_n} \ r^{-m} \ \rho^{-n} \nonumber \\
&& \qquad \qquad \qquad \times \  x^{a_1}(X) \cdots x^{a_m}(X) \ 
y^{b_1}(X) \cdots y^{b_n}(X)
\label{functions}
\end{eqnarray}
Here $W_{a_1\ldots a_m;b_1 \ldots b_n}$ is a constant and symmetric under the 
interchange of the indices, $a_1,\ldots, a_m$ and $b_1, \ldots, b_n$, 
separately. ($a_i, b_i=1,2,3$) By solving (\ref{Xxy}) $x^a$ and $y^a$ are 
expressed in terms of $X^A$ as
\begin{eqnarray}
x^i(X) &=& r \ X^{i+3}/R \ (i=1,2), \qquad x^3(X)=r \ D(X)/R, \nonumber  \\
y^a(X) &=& \rho \ X^a/D(X) \ \quad (a=1,2,3).
\label{xy}
\end{eqnarray}
The function $D(X)$ is defined by
\begin{equation}
D(X) \equiv \pm \sqrt{(X^a)^2}.
\end{equation}
Actually, these functions (\ref{xy}) correspond to the matrices 
(\ref{BaseMatrix}) in accord with the Matrix $\leftrightarrow$ Function 
correspondence (\ref{TxTy}).  
In contrast to the matrix case the coordinates $y^a$ and some of the 
functions on NC4S are ambiguous on a circle
\begin{equation}
{\cal C}= \{(X^A) \ | \ X^1=X^2=X^3=0, \ \ (X^4)^2+(X^5)^2=R^2\}.
\label{circle}
\end{equation}
The non-commutative product $\star$ is also ambiguous on ${\cal C}$.\cite{NS}

Let us introduce polar coordinates $(\theta_1,\varphi_1)$ and 
$(\theta_2,\varphi_2)$ for the two $S^2$'s; $(\theta_1,\varphi_1)$ for 
$x^a$ and $(\theta_2,\varphi_2)$ for $y^a$, respectively. By (\ref{Xxy}) 
$X^A$ can be parametrized as follows.
\begin{eqnarray}
X^1 &=& R \ \cos \theta_1 \ \sin \theta_2  \ \cos \varphi_2, 
\qquad X^2=R \ \cos \theta_1 \ \sin \theta_2 \ \sin \varphi_2, 
\nonumber \\
X^3 &=& R \ \cos \theta_1 \ \cos \theta_2, \qquad 
\qquad X^4 =R \ \sin \theta_1 \ \cos \varphi_1, 
\nonumber \\
X^5 &=& R \ \sin \theta_1 \ \sin \varphi_1
\label{S2S2}
\end{eqnarray}
Because $(\theta_1,\varphi_1,\theta_2,\varphi_2)$ and 
$(\pi-\theta_1,\varphi_1,\pi-\theta_2,\varphi_2+\pi)$ yield the same $X^A$  
the parametrization (\ref{S2S2}) describes two $S^4$'s:  
the range of $\theta_1$ must be restricted to $0 \leq \theta_1 \leq \pi/2$
for one $S^4$, and the remaining range $\pi/2\leq \theta_1 \leq \pi$ will 
correspond  to the other $S^4$. Therefore $S^2 \times S^2$ is devided into 
$S_+^2 \times S^2$ and $S_-^2 \times S^2$.  The manifold described by 
(\ref{S2S2}) is two $S^4$'s joined at the circle ${\cal C}$. We will call 
this manifold $S^4 \natural S^4$. $D(X)=R \ \cos \theta_1$ is 
positive for the first $S^4$ and negative for the second. 
To obtain a single $S^4$ we must restrict 
the manifold to the portion with $D(X) \geq 0$. In what follows we will 
assume $D(X) \geq 0$. In the case of the matrix configuration the corresponding 
matrix $\hat{D}$ (\ref{D}) was defined such that eigenvalues with both signs are 
included. Therefore we must introduce an appropriate projection matrix 
when we compute
a trace in evaluating the action integral.  
In the parametrization (\ref{S2S2}) $X^{1}$, $X^{2}$ and $X^{3}$ vanish 
at $\theta_1=\pi/2$ and the variables $\theta_2$, $\varphi_2$ become redundant.
This is the origin of the singularity of the star product and functions 
at ${\cal C}$ ($D(X)=0$). 

Since $S^4$ is embedded in the flat ${\mathbb R}^5$, the standard 
round metric 
\begin{eqnarray}
ds_4^2&=& (dX^A)^2 \nonumber \\
& =& R^2 \left( d\theta_1^2+\sin^2 \theta_1 \ d\varphi_1^2 \right)
+ R^2 \ \cos^2 \theta_1 \ \left( d\theta_2^2+\sin^2 \theta_2 
\ d\varphi_2^2 \right)
\label{S22}
\end{eqnarray}
yields the correct line element. We will equip our $S^4$ with this metric. 
The metric (\ref{S22}) in the polar coordinate 
($\theta_i$, $\varphi_i$) is also singular at $\theta_1=\pi/2$; 
the coefficients of $d\theta_2^2$ and $d\varphi_2^2$ vanish at 
$\theta_1=\pi/2$. But this singularity is an apparent singularity common 
to all the polar coordinates. Instead the functions (\ref{functions}) are
non-singular in the polar coordinate. 

\subsection{Volume forms}
\hspace*{5mm}
The volume form of the round metric in the $X^A$ coordinate system 
is given by  
\begin{equation}
d(\mbox{volume}) = \frac{R}{X^3} \ dX^1 \wedge dX^2 \wedge dX^4 \wedge dX^5.
\label{volume}
\end{equation}
Here the coordinates $X^{1,2,4,5}$ are treated as independent variables. 
This does not coincide with   
\begin{eqnarray}
d(\mbox{inv volume}) &=& \frac{1}{RD^2 X^3} \ dX^1 \wedge dX^2 \wedge 
dX^4\wedge  dX^5 \nonumber \\
&& =  (dx^1 \wedge dx^2/rx^3) \wedge (dy^1 \wedge dy^2/\rho y^3 ),
\label{invvolume}
\end{eqnarray}
the product of the $SO(3)$ invariant measures on the two $S^2$'s. These volume
forms are related by 
\begin{equation}
d(\mbox{inv volume}) = \frac{1}{R^2D(X)^2} \ d(\mbox{volume}). 
\label{VtoV}
\end{equation}
The difference of the volume forms will result in the breaking of a part of 
the cyclic property of the integration, which corresponds to the cyclic 
property of the trace of matrices. However, it is possible to write down 
the action integral for the scalar field theory. 

\subsection{Scalar field theory}
\hspace*{5mm}
As a simple example of non-commutative field theories on $S^4$ 
let us consider a scalar field theory. 
We propose the matrix action for a real scalar theory on NC4S by 
\begin{equation}
S_{\mbox{scalar}}= \frac{1}{(N/2)} \ Tr_+ \left\{\left( -\frac{1}{2} \ 
[\hat{X}^A_0, \hat{\Phi}]^2+V(\hat{\Phi})\right) \ \frac{3\hat{D}^2}{R^2}
\right\}.
\end{equation}
($N=(2j_1+1)(2j_2+1)$ ) Here $\hat{\Phi}$ is an $N$ by $N$  hermitian matrix 
representing a scalar field and $V(\hat{\Phi})$ is a potential. 
$\hat{D}$ is a matrix defined in (\ref{D}). The insertion of $\hat{D}^2$ is 
motivated by the relation (\ref{VtoV}) of the two volume forms. 
$Tr_+$ denotes a trace restricted to the subspace $\Lambda_+$ of positive 
eigenvalues of $\hat{D}$. We also denote the subspace of negative eigenvalues 
as $\Lambda_-$. The whole operation is equivalent to inserting into the trace 
of the Lagrangian a matrix $\hat{D}_+^2=\hat{D}^2 \ \hat{P}_+$, 
where $\hat{P}_+$ is a projection matrix onto $\Lambda_+$. 

One may think that this is not allowed because this will break the unitary 
symmetry $\hat{\Phi} \rightarrow \hat{U} \ \hat{\Phi} \ \hat{U}^{-1}$ of 
the action. In the case of a scalar theory the background $\hat{X}_0^A$ has 
already broken this symmetry. To construct gauge theories we need the unitary
symmetry. But $\hat{D}_+^2$ commutes with $U(N/2) \otimes U(N/2)$. 
This symmetry can be turned into the non-commutative $U(1)$ gauge symmetry. 
When we consider a $U(m)$ gauge theory, we must enlarge the matrices 
$\hat{X}^A_0$, $\hat{\Phi}$ by  tensor products with  $m$ by $m$ matrices and 
the matrix $\hat{D}_+^2$ must be chosen to be an identity in this 
extra space. 

This projection enforces the restriction to one $S^4$ of $S^4 \natural S^4$ and
must be performed once with the operation of the trace. Since the two terms 
in the action have the form $Tr_+ \ ( A \ A \ \cdots \ A ) \ \hat{D}^2 
= Tr \ ( A \ A \ \cdots \ A \ \hat{D}_+^2 )$ 
\footnote{$A=[\hat{X}^A_0, \hat{\Phi}]$ in the 
first term of the action and $A=\hat{\Phi}$ in the second.}, the location of 
$\hat{D}_+^2$ inside the trace is irrelevant due to the cyclic property of the 
trace. 

Let us work out the explicit form of the action for the case $j_1=j_2=1/2$ and
see the effect of the projection onto $\Lambda_+$. From (\ref{roundS4}) 
the values of $\alpha$ and $\beta$ are $\alpha=3R$, $\beta=3R$. 
Then the matrix configuration (\ref{matrixconfiguration}) is given by 
\begin{equation}
\hat{X}^a_0 = \frac{R}{2} \left( 
      \begin{array}{cc}
    \sigma_a & 0 \\
    0 & -\sigma_a \end{array} \right), \quad 
\hat{X}^4_0 = R \left( 
      \begin{array}{cc}
    0 & \bm{1} \\
    \bm{1} & 0 \end{array} \right), \quad
\hat{X}^5_0 =i R \left( 
      \begin{array}{cc}
    0 & -\bm{1} \\
    \bm{1} & 0 \end{array} \right). 
\end{equation}
Here $\sigma_a$ are Pauli matrices and $\bm{1}$ is a 2 by 2 identity matrix. 
$\hat{P}_+$ and $\hat{D}^2$ are  given by
\begin{equation}
\hat{P}_+ = \left( 
      \begin{array}{cc}
    \bm{1} & 0 \\
    0 & 0 \end{array} \right), \quad 
\hat{D}^2 = \frac{R^2}{4} \ \left( 
      \begin{array}{cc}
    \bm{1} & 0 \\
    0 & -\bm{1}   \end{array} \right).
\end{equation}
As for $\hat{\Phi}$ we put  
\begin{equation}
\hat{\Phi} = \left( 
      \begin{array}{cc}
    A & B \\
    B^{\dagger} & C \end{array} \right),
\end{equation}
where $A$, $C$ are 2 by 2 hermitian matrices and $B$ a 2 by 2 matrix.  
By simple calculation we obtain the result.
\begin{eqnarray}
-Tr_+ \ [\hat{X}_0^a, \hat{\Phi}]^2 \ \hat{D}^2 &=& \frac{R^4}{16} \ 
tr \{-  [\sigma_a, A]^2 +   (\sigma_a B+ B \sigma_a) \ 
(\sigma_a B+ B \sigma_a)^{\dagger}\}, \nonumber \\
-Tr_+ \ [\hat{X}_0^4, \hat{\Phi}]^2 \ \hat{D}^2 &=& \frac{R^4}{4} \ tr \{ 
-(B-B^{\dagger})^2 +(A-C)^2 \}, \nonumber \\
-Tr_+ \ [\hat{X}_0^5, \hat{\Phi}]^2 \ \hat{D}^2 &=& \frac{R^4}{4} \ tr \{ 
(B+B^{\dagger})^2 +(A-C)^2 \}.
\end{eqnarray}
Here $tr$ is the trace of 2 by 2 matrices.  Therefore the kinetic part of the 
action $\frac{3R^2}{64} tr \{ -  [\sigma_a, A]^2+   (\sigma_a B+ B \sigma_a) \ 
(\sigma_a B+ B \sigma_a)^{\dagger}+ 16 B B^{\dagger} + 8(A-C)^2 \}$ yields a 
necessary damping factor for all the variables in the path integral. 
$A$ and $C$ may be regarded as matrices associated with $S^4$'s with 
$D(X) \geq 0$ and $D(X) \leq 0$, respectively, while $B$ a matrix connecting 
the two $S^4$'s. The above example shows that only the matrix $A$ has the \lq
kinetic term', {\it i.e.} the commutator term. $B$ and $C$ are auxiliarly 
fields. 

According to the Matrix $\leftrightarrow$ Function correspondence, this action
can be rewritten in terms of the non-commutative product.  The corresponding
scalar field on $S^4$ is denoted by $\Phi(X)$. We define the non-commutative 
derivative $\nabla^A$ by the commutator. 
\begin{equation}
\nabla^A \Phi(X) \equiv -i [X^A, \Phi(X)]_{\star}
\end{equation}
Here the star-commutator $[ \cdot  , \cdot ]_{\star}$ is defined by 
$[F,G]_{\star}=F \star G- G \star F$. In the Matrix $\leftrightarrow$ 
Function correspondence the restricted trace $Tr_+$ is replaced by the volume 
integration only on one $S^4$ with $D \geq 0$, {\em i.e.}, 
$S^2_+ \times S^2$.  
\begin{equation}
\frac{1}{(N/2)} \ Tr_+ \rightarrow \frac{1}{\mbox{Vol}(S_+^2 \times S^2)} 
\ \int_{S^2_+ \times S^2} d(\mbox{inv volume})
\label{integral}
\end{equation}
We must use the invariant volume form and it is given by (\ref{invvolume}). 
The prefactor of the integral is the inverse of the volume of half  
$S^2 \times S^2$, 
\begin{equation}
\mbox{Vol}(S_+^2 \times S^2)= \frac{1}{2} \times (4 \pi)^2 = 8 \pi^2.  
\label{Vol}
\end{equation}
The action integral is now given by
\begin{eqnarray}
&& S_{\mbox{scalar}}  \nonumber \\ && =  \ \int_{S_+^2 \times S^2} 
\frac{d(\mbox{inv volume})}{\mbox{Vol}(S_+^2 \times S^2)} 
 \ \left\{
\frac{1}{2} \ (\nabla^A \  \Phi(X))\star (\nabla^A \  \Phi(X)) 
+ V(\Phi)_{\star} \right\} \star \frac{3D^2}{R^2}. \nonumber \\ &&
\label{scalar2}
\end{eqnarray}
Here $V(\Phi)_{\star}$ is obtained from $V(\Phi)$ by replacing the ordinary 
products by $\star$. 
By using eq (\ref{VtoV}) we replace the integration 
measure by that of $S^4$ equipped with the round metric. 
\begin{equation}
S_{\mbox{scalar}}=  \ \int_{S^4} 
\frac{d(\mbox{volume})}{ D^2 \ \mbox{Vol}(S^4)} \ \left\{
\frac{1}{2} \ (\nabla^A \  \Phi(X))\star (\nabla^A \  \Phi(X)) 
+ V(\Phi)_{\star} \right\} \star D^2.
\label{scalar3}
\end{equation}
Here Vol($S^4$) is the volume of $S^4$.
\begin{equation}
\mbox{Vol}(S^4)= \frac{8\pi^2}{3} \ R^4
\end{equation}
Let us notice that $D^2$ in the denominator of the measure and $ D^2$
in the integrand do not cancel. 

Now let us consider the commutative limit ($j_1, j_2 \rightarrow \infty$, 
$R$ fixed) In this limit we expect that the action (\ref{scalar3}) will 
reduce to 
\begin{equation}
S_{\mbox{comm}}=  \int_{S^4} 
\frac{d(\mbox{volume})}{\mbox{Vol}(S^4)} \ \left\{
\frac{1}{2} \ G^{AB}(X) \ \partial_A \   \Phi(X)\  \partial_B  \Phi(X) 
+ V(\Phi) \right\}.
\label{comm}
\end{equation}
Here $G^{AB}$ $(A,B=1,\ldots,5)$ and its inverse,$G_{AB}$, 
are the metric in the tangent space $T_p (S^4)$ and cotangent space $T^{\ast}_p(S^4)$ at $p=(X^A)$ defined by $ds_4^2=G_{AB} \ dX^A \ dX^B$, 
$G_{AB} \ G^{BC}=\delta_{AC}-X^A X^C/R^2$, $X^A \ G_{AB}=X^A \ G^{AB}=0$; 
$G^{AB}$ is given by 
\begin{eqnarray}
G^{ab} &=&  \delta^{ab}- \frac{1}{R^2} \ X^a \ X^b, \nonumber \\
 G^{ai} &=& G^{ia} = - \frac{1}{R^2} \ X^a \ X^i, \nonumber \\
G^{ij} &=&  \delta^{ij} - \frac{1}{R^2} \ X^i \ X^j 
\end{eqnarray}
We have not proved this yet. The proof will depend on the explicit form of the
star product and the derivative $\nabla^A$ for general values of $j_1$, $j_2$. 
We hope to report on this in the future. 

In the commutative limit the scalar field $\Phi(X)$ has the form 
(\ref{functions}) with the summations extending to infinity. This field 
is singular on the circle, $D(X)=0$, and differentiation may produce stronger
singularities. Is the action (\ref{comm}) well-defined? 
We will show that this integral is finite. 
Let us note that all the terms in $G^{ab}$ and $G^{ai}$ 
are proportional to $X^a$, $X^b$ except for the term 
$\delta^{ab}$ in $G^{ab}$. Therefore in the kinetic term in 
(\ref{comm}) the derivatives with respect to $X^a$  appear in the 
combination $X^a \ \partial_a$. Since the singulatity of $\Phi$ is at most 
the ambiguity $X^c /D(X)$ and this satisfies 
$X^a \ \partial_a \ (X^c/D)=0$, this type of differentiation does not create 
stronger singularities. 
The term in $G^{ab}$ proportional to $\delta^{ab}$ will create a singularity 
$\delta^{ab} \ \partial_a \ (X^c/D)\ \partial_b \ (X^d/D) = 1/D^2 \ 
(\delta^{cd}-X^c \ X^d /D^2)  $ but this singularity is integrable. 
Therefore the commutative limit of the non-commutative 
scalar field theory is well-defined.

Finally, we must comment on the cyclic property of the integral 
(\ref{integral}). Usually, the trace of matrices $\hat{A}$, $\hat{B}$ 
satisfies the cyclic property $Tr \hat{A}\hat{B}=Tr \hat{B}\hat{A}$ and 
by the Matrix $\leftrightarrow$ Function correspondence the integral of 
the functions $A(X)$, $B(X)$ satisfies the similar property 
$\int dX A(X) \star B(X) = \int dX B(X) \star A(X)$. Our trace $Tr_+$, 
however, does not have this property: 
$Tr_+ \ \hat{A} \hat{B} \hat{D}^2 \equiv Tr \hat{A} \hat{B} \hat{D}_+^2
\neq Tr_+ \ \hat{B} \hat{A}\hat{D}^2$. 
The corresponding integral does not have this propery, either. 
By the same reason as for the trace, $Tr_+$, this is 
not an obstacle for writing the action integral for the scalar theory. 
When we consider correlation functions, however, we must keep the ordering, 
not just the cyclic ordering, in the correspondence.
\begin{equation}
\frac{1}{(N/2)} \ Tr_+ \ \hat{A}_1 \hat{A}_2 \cdots \hat{A}_n \hat{D}^2
\leftrightarrow 
 \int_{S^4}\frac{d(\mbox{volume})}{D^2\mbox{Vol}(S^4)} \  A_1(X) \star 
A_2(X) \star \cdots \star A_n(X) \star D^2
\end{equation}

When we consider the gauge theories we need the cyclic property to implement 
the gauge symmetry. On the Matrix theory side we have $U(N/2) \times U(N/2)$ 
symmetry as mentioned above. If $\hat{H}$ is a matrix of the form 
\[
\hat{H} = \left( 
      \begin{array}{cc}
    H_+ & 0 \\
    0 & H_- \end{array} \right),
\]
where the first block is for $\Lambda_+$ and the second for $\Lambda_-$, 
we have $Tr_+ \ \hat{A} \hat{H}\hat{D}^2 = Tr_+ \ \hat{H} \hat{A}\hat{D}^2$ 
for any matrix $\hat{A}$.  The matrix $\hat{H}$ keeps the subspaces 
$\Lambda_{+}$, $\Lambda_{-}$  invariant separately. On the function side, 
since $D(X)$ is nothing but $x^3$ as one can see from (\ref{xy}), 
the functions which correspond to the matrix $\hat{H}$ are $x^3$ and 
$y^{1,2,3}$. Then for an arbitrary function $A(X)$ on $S^4$ and a function 
$H(X) \equiv H(x^3(X),y^a(X))$ which is a polynomial of $x^3(X)$ and $y^a(X)$, 
we can check the following formula by using (\ref{starS2}) and (\ref{starS4}). 
\begin{equation}
\int_{S^4}\frac{d(\mbox{volume})}{D^2\mbox{Vol}(S^4)} \  A(X) \star H(X) 
\star D^2 
= \int_{S^4}\frac{d(\mbox{volume})}{D^2\mbox{Vol}(S^4)} \  H(X)\star A(X)\star D^2
\end{equation}
Therefore we can construct a gauge theory action which is invariant under 
restricted gauge transformations generated by gauge functions which are 
polynomials of $x^3(X)$ and $y^a(X)$. The remaining gauge symmetry will  
recover in the commutative limit. The problem of the gauge symmetry needs 
further investigation.

\section{Summary}
\hspace*{5mm}
In this paper we have shown that any matrix of size $N=(2j_1+1)(2j_2+1)$ can 
be expressed in terms of the matrix $S^4$ configuration $\hat{X}^A_0$.  
Therefore in the Matrix theory version of the non-commutative field theories 
on $S^4$ the fluctuations of NC4S can be expanded in terms of $\hat{X}^A_0$. 

Then we constructed a non-commutative scalar field theory on $S^4$.  
The novel point is the projection matrix inside the trace which restricts the 
space of matrices to the subspace corresponding to one of the two $S^4$'s.  
In the description of the non-commutative field theory in terms of the star 
product this restriction corresponds to $D(X) \geq 0$. The restriction on 
the cyclic property of the trace and the integration is also discussed. 

We are now investigating the large radius limit of the non-commutative field 
theories on $S^4$. We found that Moyal deformation of ${\mathbb R}^4$ can be 
obtained in the vicinity of a point on $S^4$ in a suitable 
$j_1, j_2 \rightarrow \infty$ limit. 
Some insight into the geometry of NC4S may be obtained by studying this limit. 
The result will be reported elsewhere.\cite{NS2}
We considered only the scalar field theory in the present paper but the matrix 
formulation of the gauge theory on the NC4S is also possible.\cite{NS} 
One of the methods for the study of the matrix version of the non-commutative 
field theories, including scalar theory and gauge theory, is the numerical 
simulation.\cite{Nishimura} We are planning to perform the numerical analysis 
of non-commutative gauge theories on $S^4$. Inclusion of fermionic fields on 
NC4S is also an interesting problem. The construction of the fermion action 
on $S^4$  and the extension to the non-commutative description will be 
explored. Finally, the extension of the present work to higher dimensional 
non-commutative $S^{2n}$ will be straightforward and interesting.

\section*{Acknowledgments}
\hspace{5mm}
The work of R.N. is supported in part by Grant-in-Aid (No.13135201) 
from the Ministry of Education, Science, Sports and Culture of Japan 
(Priority Area of Research (2)).

\end{document}